\documentclass[11pt]{article}
\usepackage[dvips,dvipdf]{graphicx}
\usepackage{color}
\usepackage{amssymb} 
\newcommand{\beq}{\begin{eqnarray}}
\newcommand{\eeq}{\end{eqnarray}}
\newcommand{\be}{\begin{equation}}
\newcommand{\ee}{\end{equation}}

\newcommand \VEV [1] {\left\langle{#1}\right\rangle}

\title{The Nuclear Yukawa Model on a Lattice}
{\author{F. de Soto$^{a}$, J.C. Angl\`es d'Auriac$^b$, J. Carbonell$^b$\\
{\small\em $^a$Dpto. Sistemas F\'{\i}sicos, Qu\'{\i}micos y
Naturales; U. Pablo de Olavide, 41013 Sevilla, Spain}
\\{\small\em $^b$Laboratoire de Physique Subatomique et Cosmologie,
53 avenue des Martyrs, 38026 Grenoble, France}}}

\begin{document}
\maketitle
\bibliographystyle{unsrt}

\begin{abstract}
We present  the results of  the quantum field theory approach to nuclear Yukawa model 
obtained by standard lattice techniques.
We have considered the simplest case of two identical
fermions interacting  via a scalar meson exchange. 
Calculations have been performed using Wilson fermions in the quenched approximation.
We found the existence of a critical coupling constant above which the model
cannot be numerically solved.
The range of the accessible coupling constants is below the 
threshold value for producing two-body bound states.
Two-body scattering lengths have been obtained and compared to the non relativistic results.
\end{abstract}

\section{Introduction}\label{intr}

The application of lattice techniques to nuclear physics
is nowadays an active and fruitful  field of investigation. 
This activity has been extensively developed in the last ten years and  covers different aspects 
of the problem. 

The first one aims to obtain relevant nuclear properties from ab-initio lattice QCD calculations 
(LQCD). In this approach, the elementary fields are quarks and gluons
and all the numerical simulations depend on the very few QCD parameters: the bare coupling 
constant and the quark masses. 
The first task there is to generate the nucleon from its elementary constituents and thus 
this approach is necessarily limited
to very light nuclei. First unquenched results on NN scattering length \cite{BBOS_PRL97_06}  
and NN potentials \cite{VNN_PRL99_2007,Ishii_Lattice_09} 
have been obtained although still for large pion masses. 
Very recently,  the binding energy of the A=3 and A=4 nuclei have been computed 
\cite{Yamazaki_PRD81_2010}. 
Although being performed in the quenched approximation, this is a result that 
seemed out of  range just few years ago.
 
A second approach has been developed using the same techniques
but in the framework of effective field theories, {\it i.e.} using nucleon and mesons as elementary fields.
Lattice effective field theory (LEFT) has been first applied to study nuclear \cite{LEFT_Nuclear_matter} and neutron matter \cite{LEFT_neutron_matter} 
and was latter adapted to light nuclei \cite{LEFT_EPJA45_2010}.
The action describes non relativistic nucleons interacting via realistic, chiral inspired, NN potentials. These
potentials depend on a large number of parameters depending on the order of chiral PT  and
have proved to be very successful in the non relativistic Faddeev-Yakubovski description of the A=2,3,4 nucleon problem. 
LEFT is a quantum mechanical  description of a many body system, equivalent to a ladder potential models, that take
the simplicity of the lattice techniques to go beyond few-body methods.
 At present this allows to treat bound states of nuclei up to A=12 nucleons and can be extended well beyond.
 A review on this approach can be found in \cite{LEFT_Review_Lee_2009}.

Our aim in this work was to consider the simplest Quantum Field Theory (QFT) model of interacting
fermions which was at the origin of nuclear forces:  the Yukawa model.
Since Yukawa pioneer work \cite{Y_35}, the meson-exchange interactions  constitute the starting point for
building the NN potentials \cite{NIJ_PRC_93,AV18_PRC_95,M_PRC63_01} which, inserted in
Schrodinger-like equations, provides an "ab-initio" description of light nuclei up to $A\sim10$ \cite{LightN}.
The potential approach, however, takes into account only a small, though
infinite, fraction of diagrams of the perturbative series -- the ladder sum. 
This represents a
severe restriction of the interaction, specially taking into account
the large values of the coupling constants involved.  Chiral inspired
NN models \cite{W_NPB363_91,ORVK_PRC53_96,EGM_NPA671_00}, which can be
formally distinguished from the traditional meson-exchange ones, suffer from the same restrictions.

To incorporate the full content of the meson-exchange Lagrangian 
we have used the standard lattice techniques \cite{Montvay}, developed in the context of QCD. 
They are based on a discrete Feynman path
integral formulation of QFT and provide nowadays a genuine way to solve
non perturbatively such problems.  Preliminary results of this work can be found 
in \cite{Feli_NPB164_2007,Feli_EPJA31_2007, Feli_NPA790_2007}.
A similar study was undertaken in \cite{NT_PRL77_96} in the frame of a 
purely scalar $\phi^2\chi$ model.

A few lattice investigations of the Yukawa model with an additional
$\lambda\phi^4$ term have also been performed some time ago in the spontaneously 
broken phase \cite{Yukawa}.
Similar models were studied in the past (see for instance ~\cite{Montvay} and references therein).
and the existence of numerical instabilities  beyond a critical value of the coupling  were
found. Different schemes were used for the discretization of the meson field, obtaining critical values
of the lattice coupling slightly different, but corresponding to the same continuum limit.
More recently the phase structure of a chirally invariant lattice Higgs-Yukawa model
was studied  to establish Higgs boson mass bounds \cite{GJ_07_2,GJ_JHEP1004_2010}. 
None of these studies was however interested to investigate the existence of two-fermion bound states
in the original Yukawa model.

The plan of the paper is as follows. In section \ref{The_model} we describe the model and its
discretization as well as the approximations in use.  Section \ref{Monte_Carlo} contains
the detail of the Monte Carlo simulations and the results for the renormalized fermion mass. The
two-body binding energies and scattering lengths are presented in section \ref{Two_Body} and compared
to the non relativistic results. Some concluding remarks are finally drawn in section 
\ref{Conclusion}.

\section{The model}\label{The_model}

We have considered the simplest renormalizable Quantum Field Theory  describing a  fermion ($\Psi$) -- meson ($\Phi$)
interaction. It is given  by the Lagrangian  density:
\begin{equation}\label{lagrangian}
\mathcal{L}=\mathcal{L}_{D}(\bar\Psi,\Psi) + \mathcal{L}_{KG}(\Phi) + \mathcal{L}_{I}(\bar\Psi,\Psi,\Phi)\ ,
\end{equation}
with, in Euclidean space, 
\beq
\mathcal{L}_{D}(\bar\Psi,\Psi)      &=&  \overline\Psi\left(\partial_\mu\gamma^\mu + m_0\right) \Psi\ , \label{LD}\\
\mathcal{L}_{KG}(\Phi)              &=&  \frac{1}{2}\left(\partial_\mu\Phi\partial^\mu\Phi + \mu_0^2\Phi^2 \right) \ ,\label{LKG} \\
\mathcal{L}_{I}(\bar\Psi,\Psi,\Phi) &=&  g_0 \overline\Psi\Gamma\Phi\Psi + \lambda_0 \Phi^4\ , \label{L_I}
\eeq
The fermion field is supposed to describe a nucleon (N) and the meson field a -- more or less
fictitious -- scalar particle ($\sigma$) responsible for the attractive part of the NN potentials. 
The Lagrangian depends on four parameters: the fermion $m_0$ and  meson $\mu_0$ masses and two
dimensionless coupling  constants $g_0$ and  $\lambda_0$.  The Yukawa coupling in Eq.  (\ref{L_I})
admits several possibilities depending on the choice of the $\Gamma$ matrix:  $\Gamma=I$ for the 
scalar case and $\Gamma=i\gamma_5$ for the pseudo-scalar one. In this work we will restrict  to the
scalar coupling, for it is known to produce a stronger fermion-fermion attraction.

This theory can be treated perturbatively by computing order by order the contributions
in $g_0$, as is done for example in QED for computing the anomalous magnetic moment of the
electron. Bound states nevertheless  appear only non-perturbatively, when contributions at all orders in the
coupling are taken into account. 

Non-perturbative tools  are extensively used in the context of QCD. Among them, the lattice 
techniques provide a reliable numerical method for solving any QFT on a  discretized Euclidean
space-time. A very basic description of these techniques  is given below. The interested reader can
found a detailed explanation  in e.g. \cite{Montvay}.

The vacuum expectation values of the operators involved are obtained in the Path Integral approach which consist  in computing  integrals like:
\begin{equation}\label{vev}
\VEV{\mathbf{O}(\bar\Psi,\Psi,\Phi)}\ =\ \frac{1}{Z} \; \int[d\bar\Psi][d\Psi][d\Phi] \mathbf{O}(\bar\Psi,\Psi,\Phi) \; e^{-S_E[\bar\Psi,\Psi,\Phi]}\ ,
\end{equation}
where the Euclidean action plays the role of a probability distribution in a Monte Carlo simulation. 

The discretized Euclidean action $S_E$ can be written, according to Eq.  (\ref{lagrangian}), 
in the form:
\[ S_E=a^4\sum_x \mathcal{L} = S_D + S_{KG} + S_I\]
where $a$ is the lattice spacing and $x$ denotes a point with coordinates 
$x_{\mu}=a n_{\mu}$  ($\mu=1,2,3,4$) and  $n_{\mu}=1\ldots, L_{\mu}$.
In practice we have taken equal spatial dimensions $L_1=L_2=L_3=L$ and  
a temporal one with $L_4=2L$. 

For the  free Dirac action $S_D$, we have used Wilson fermions. 
They consist in adding to the, naively discretized,  derivative term of the Dirac Lagrangian (\ref{LD})   a
Laplacian operator in order to remove the spurious poles at the boundaries of the Brillouin zone \cite{Montvay}. 
The free Dirac  action is  then written as a bilinear form in the dimensionless fermion fields $\psi= \sqrt{a^3\over2\kappa} \Psi$:
\beq\label{S_D}
S_D = \sum_{xy} \bar\psi_x D^W_{xy} \psi_y
\eeq
where
\beq\label{DiracWilson}
D^W_{xy} = \delta_{x,y} - \kappa \sum_\mu \left[ \left(1-\gamma_\mu\right) \delta_{x,y-\mu} + \left(1+\gamma_\mu\right) \delta_{x,y+\mu} \right]
\eeq
is the Dirac-Wilson operator,   and
\beq\label{kappa}
\kappa=\frac{1}{8+2am_0}
\eeq 
is the hopping parameter.

In terms of the dimensionless meson field $\phi=a\Phi$, the discrete Klein-Gordon action reads:
\beq\label{S_KG}
S_{KG} = \frac{1}{2} \sum_{x} \left[ \left( 8 + a^2\mu_0^2 \right) \phi_x^2 - 2 \sum_{\mu} \phi_{x+\mu}\phi_x \right]
\eeq

The  interaction term takes the form
\beq\label{S_I}
S_{I} = g_L \sum_{x} \bar\psi_{x}\phi_{x}\psi_{x} + \lambda_0 \sum_{x} \phi_{x}^4
\eeq
where $g_L=2\kappa g_0$ is the  lattice Yukawa coupling. 

Taking together the fermion (\ref{S_D}), meson (\ref{S_KG}) and interaction  (\ref{S_I}) terms, the lattice Euclidean action can be written in the form:
\beq	
S_E(\bar\psi,\psi,\phi) =  \sum_{xy} \bar\psi_x D_{xy} \psi_y + S_M(\phi)
\eeq
where  $S_M$ includes both the Klein-Gordon and the $\lambda \phi^4$ term 
\beq\label{S_M}   
S_M= S_{KG} +    \lambda_0 \sum_{x} \phi_{x}^4  
\eeq
and the fermionic part is written in terms of  the interacting Dirac operator
\begin{equation}\label{DO}
D=D^W + g_L \phi
\end{equation}

Notice that the model is now made dimensionless. When needed, the physical quantities -- masses, energies, etc. -- will be
given in terms of the lattice spacing.

One of the most demanding issues when computing the vacuum expectation values 
(\ref{vev}) comes from  the Grassmannian character of fermion fields, that have to be
integrated out by algebraic methods. For example the  fermion propagator, corresponding to
$\mathbf{O}(\bar\psi,\psi,\phi)=\psi_x\overline\psi_y$,   
\begin{equation}\label{unquenched}
S(x,y)\ =\ \VEV{\psi_x\overline\psi_y}\ =\ \frac{1}{Z} \int[d\phi]\ D^{-1}_{xy}\ {\rm det} [D(\phi)]\ e^{-S_{M}(\phi)}\ ,
\end{equation}
implies the evaluation of a determinant and inverse of an operator that, even for
moderate lattices,  $V\sim 24^4$, has a dimension of $\sim 10^6$. Moreover, if a Monte
Carlo simulation is to be done using  Eq. (\ref{unquenched}), the probability distribution
for meson configurations is given by  $e^{-S_M(\phi)-\log(\det(D))}$, what means
evaluating a large determinant in every Monte Carlo step . This can be avoided by the
use of Hybrid Monte Carlo techniques that nevertheless are the main source of time spent
in the simulation. This task is considerably  simplified in the ``quenched'' approximation
that, from  the computational point of view consists in setting $\rm{det(D)}$ independent
of  the meson field in the fermionic integral.

\begin{figure}[h!]
\vspace{-0.5cm}
\begin{center}
\includegraphics[width=7.cm]{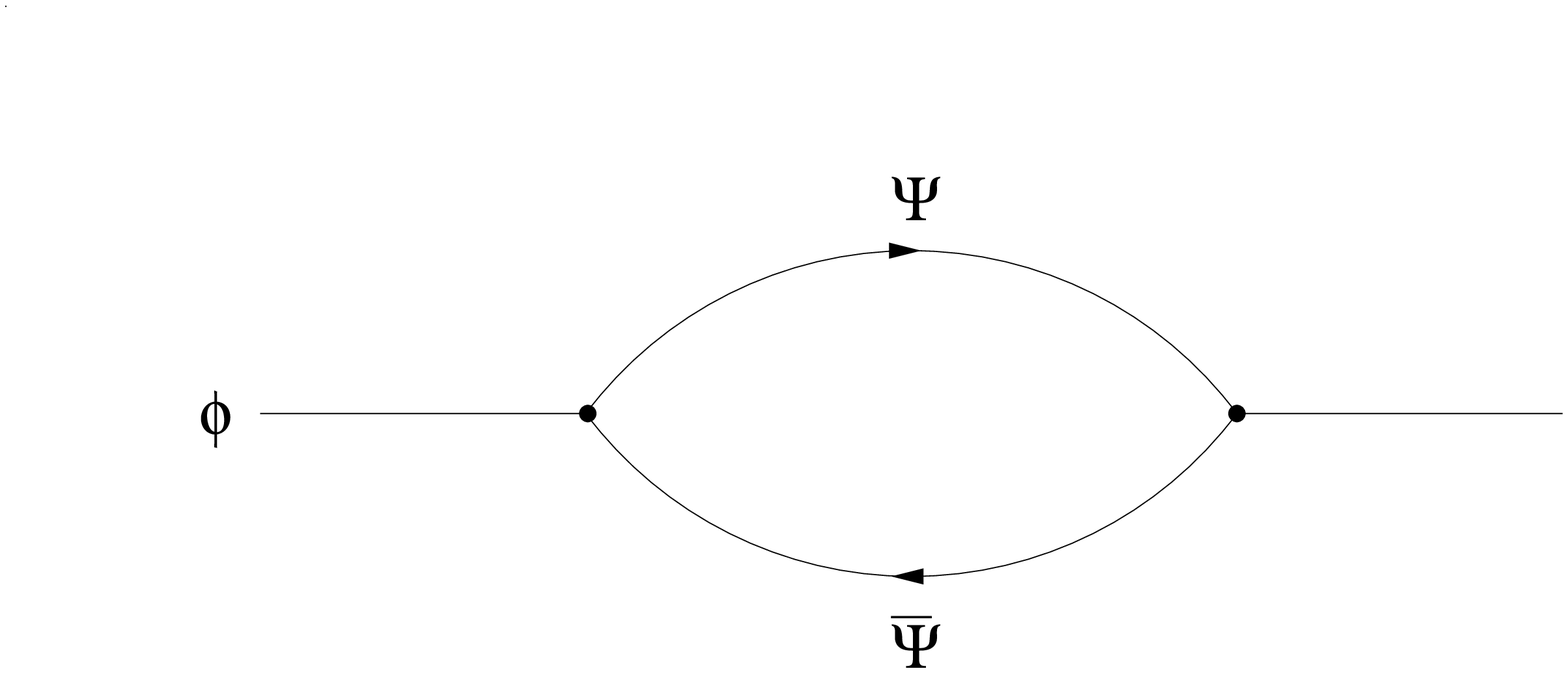}
\end{center}
\vspace{-0.5cm}
\vspace{.cm} \caption{The quenched approximation neglects the possibility 
for a meson $\Phi$ to create a virtual fermion-antifermion pair $\Psi\bar\Psi$.} \label{quench}
\end{figure}

From a physical point of view, the quenched approximation avoids the possibility for a meson to create a
virtual  nucleon-antinucleon pair $\phi\to\bar\psi\psi$ (see figure \ref{quench}). 
Due to the heaviness of the nucleon with respect to the exchanged meson  this
approximation is fully justified in low energy nuclear  physics  and implicitly assumed  in all  the potential models.

We have furthermore chosen in our simulation to neglect the meson self-interaction term setting $\lambda_0=0$ in (\ref{S_M}).
This choice is consistent only in the quenched approximation. 
In a full QFT treatment of this model, the fermions loops will  generate meson 
self-interactions, that require a $\lambda_0\phi^4$ counter-term for renormalizability.

The model  depends on three dimensionless 
parameters  $g_0$, $a\mu_0$, and $am_0$  usually  set via the hopping parameter (\ref{kappa}).  
These parameters appearing in the Lagrangian are not physical:  they are 
modified by the interaction so that they have to be renormalized. 
Our first task  to map the bare quantities into the renormalized parameter space
\[(a\mu_0,\ g_0,\ am_0)\ \ \Rightarrow\ \ (a\mu_R,\ g_R,\ am_R) \]

In the quenched approximation and for $\lambda_0=0$, mesons do not interact each other
and therefore its mass renormalize trivially $a\mu_R=a\mu_0$. 
In the following we will omit subscripts and denote the dimensionless meson mass by  $a\mu$. 
The lattice spacing $a$ does never appear explicitly and it has to be fixed 
by setting a physical dimensional quantity. We do that by identifying the unchanged meson
mass $\mu$ to a physical meson of $\mu=0.65$ GeV, a typical value used in the NN models. 
If we are using in our simulations $a\mu=0.1$, the lattice spacing is given by $a={a\mu\over\mu}=\frac{0.1}{0.65\; {\rm GeV}}\approx 0.15 \;{\rm
GeV}^{-1} \approx 0.03 \;{\rm fm}$. 

Coupling constant renormalization is a more involved issue. 
Renormalized coupling constants were  computed in a previous work \cite{Feli_NPB164_2007}.
For the scalar coupling  no sizable effect of renomalization was found, i.e. $g_R\approx g_0$ in a wide range of momenta.
From now on, we will denote by $g$ this unique coupling constant. 

The remaining task for finding the adequate parameter space  is thus  to compute the renormalized fermion mass
as a function of the bare parameters.
This will be described  in the next section.

\section{Monte-Carlo simulation}\label{Monte_Carlo}

\subsection{Generating field configuration}

In the above defined conditions, the generation of meson field configurations is  straightforward.
This can be seen by writing the Klein-Gordon action  (\ref{S_KG}) in Fourier space
\beq\label{S_KG_k}
S_{KG} =  \sum_{k} \frac{\left|\tilde\phi_k\right|^2}{2\sigma_k^2}  \ ,
\eeq
where $\tilde{\phi}_k $ denotes the Fourier transform of the meson field $\phi_x$
\[\tilde{\phi}_k ={1\over \sqrt{V}} \sum_x \phi_x e^{-ik\cdot x} \]
and we have introduced the quantity
\[  \sigma_k^2=\frac{1}{\hat{k}^2+a^2\mu^2}\]
which depends on the lattice discretized momenta 
\[ \hat{k}_\mu = 2 \sin{k_\mu \over 2}   \qquad\quad  k_{\mu}={2\pi\over L_{\mu}}\; n_{\mu}\]
One can see from equation (\ref{S_KG_k}) that the different Fourier modes $\tilde{\phi}_k$ 
of the meson field appearing in the action are decoupled and can be generated independently.

The Monte Carlo algorithm becomes then 
trivial as it is enough to produce at each lattice point $k$, independent  complex scalar 
fields with a  probability density given by
\begin{equation}\label{tirage}
P( \tilde\phi_k ) \sim \exp\left[- \frac{1}{2}{ |\tilde\phi_k|^2\over \sigma^2_k}\right] = 
   \exp\left[- \frac{1}{2}{ {\rm Re}[\tilde\phi_k^2] \over \sigma^2_k}\right]
   \exp\left[- \frac{1}{2}{ {\rm Im}[\tilde\phi_k^2] \over \sigma^2_k}\right]
\end{equation}
i.e.  centered Gaussian distributions,  both for their real and imaginary
parts, with a variance $\sigma_k$ depending on $k$ and the constraint imposed by the reality of
$\phi_x$. This method generates configurations that are statistically independent, thus saving a large
amount of computing time with respect to the Metropolis algorithm. 

The scalar fields in configuration space $\phi_x$ are finally obtained by performing an inverse
Fourier transform on $\tilde{\phi}_k $. It follows from the particular form  (\ref{tirage}) 
that $\phi_x$  are also centered Gaussian with a width $\sigma$, independent of $x$, given by
\begin{equation}\label{sigmax}
\sigma^2(L,a\mu) = {1\over V} \sum_k \frac{1}{\hat{k}^2+a^2\mu^2}  
\label{eq:sigma2}
\end{equation}
Note however that the $\phi_x$ are now correlated. The correlation function is given by
\beq\label{latticeV} 
\Delta(x-y) = < \phi_x\phi_y>= \sum_k  \frac{1}{ \hat{k}^2 +a^2\mu^2 }  e^{ik(x-y)}
\eeq
which is the scalar propagator in configuration space. 
It is interesting to note that the parameter $\sigma$ in Eq. (\ref{sigmax}) is related
to the lattice regularization of the potential at the origin,
\beq
V(0) = - g^2 \Delta(0) = -g^2 V \sigma^2 \ .
\eeq

\begin{figure}[h!]
\begin{center}
\begin{minipage}[h!]{8cm}
\includegraphics[width=8cm]{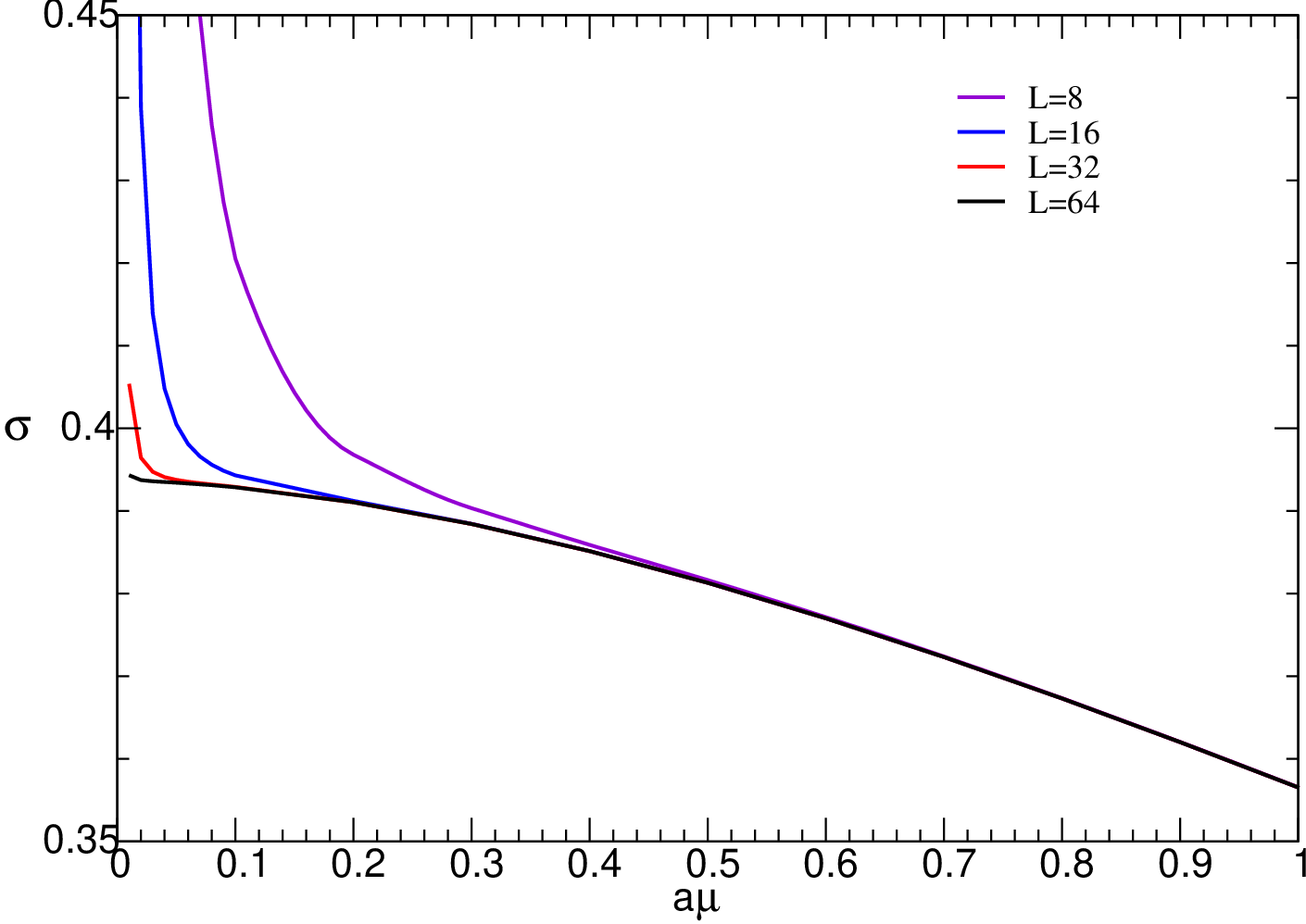}
\vspace*{0.1cm}
\caption{$a\mu$-dependence of the variance for the meson fields in configuration space.}
\label{sigma_mu}
\end{minipage}
\hspace{0.5cm}
\begin{minipage}{8cm}
\includegraphics[width=7.5cm]{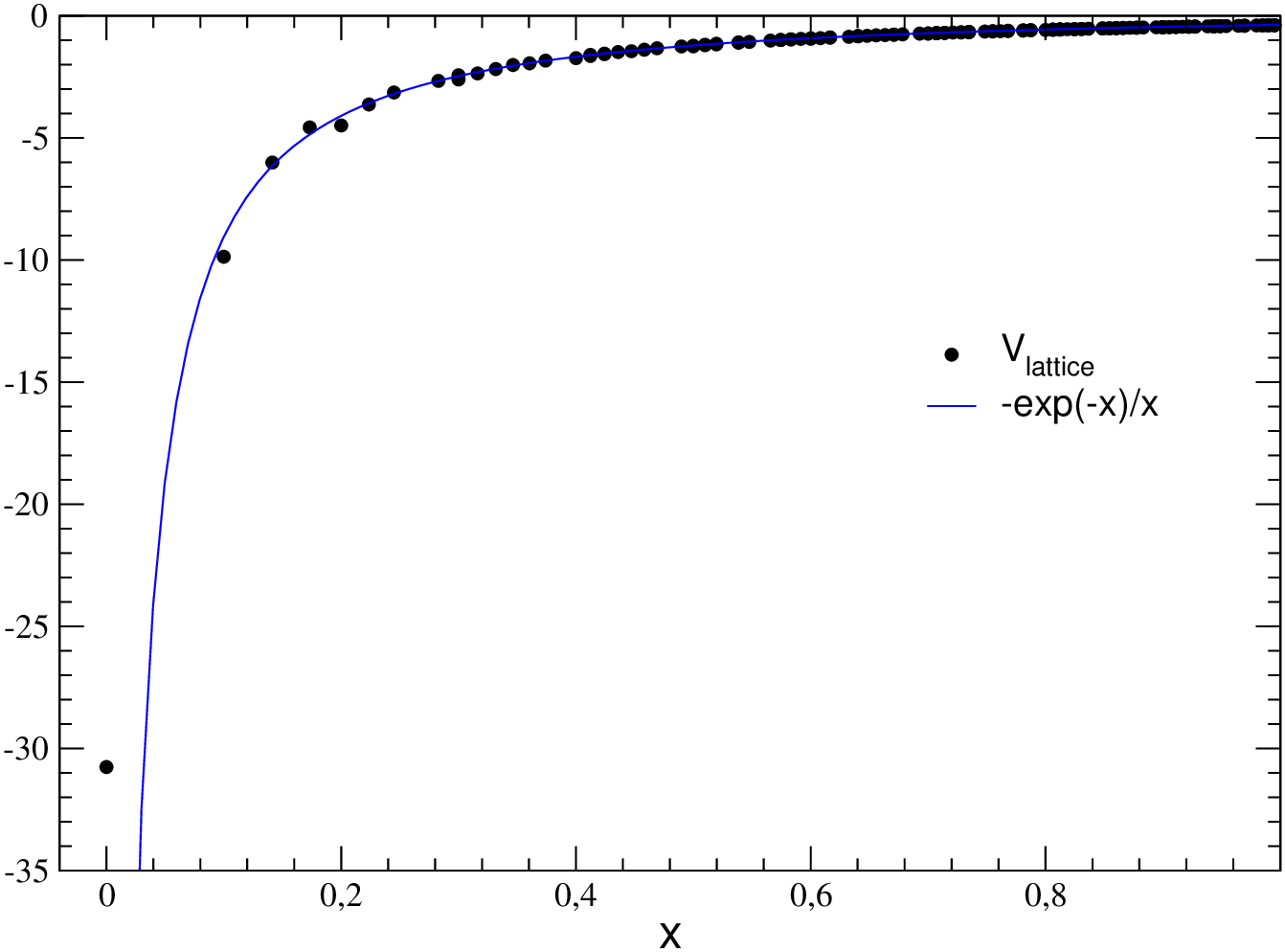}
\vspace*{0.1cm}
\caption{Discrete Yukawa potential ($V_{Lattice}$) for $L=48$ and $a\mu=0.1$ (Black dots) 
compared to the continuum one (Solid line).}
\label{latticepotential}
\end{minipage}
\end{center}
\end{figure}

The $a\mu$-dependence of $\sigma$ is displayed in figure  \ref{sigma_mu}  for different
values of the lattice size, $L$. It behaves  like $1/a\mu$  in the two trivial limits $a\mu\to 0$
and $a\mu\gg1$ with a plateau in between, which is the region  we are interested in. 
The parameter $\sigma$  allows a discussion of the discretization and finite volume errors
in terms of physically well defined limiting cases:
\begin{itemize}
\item For large values of $a\mu$, the $\hat{k}^2$ term in (\ref{sigmax}) becomes negligible and 
 $\sigma={1\over{a\mu}}$ for any value of $L$. This correspond to a contact 
 interaction between fermions.  
\item In the limit  $a\mu\to 0$  the sum (\ref{sigmax}) is dominated by the mode
 $\tilde\phi_{k=0}$ which  generates a behavior \[\sigma^2={1\over V}{1\over{a^2\mu^2}}+\ldots\]
This limit corresponds to the mean field approach of the problem.
\end{itemize}

\vspace{0.8cm}

The non-relativistic Yukawa potential can be computed for the 
discrete lattice using Eq. (\ref{latticeV}) but summing only over spatial 
directions. The resulting potential ($V_{Lattice}$) for $a\mu=0.1$ and $L=48$ 
is represented in figure \ref{latticepotential} (Black dots) and compared to 
the continuum result (Solid line). The main effect of discretization is the 
regularization at the origin and is seen to be negligible  
beyond the very first points. This potential will be used in section \ref{Two_Body} 
to compare continuum and lattice results.

The appearance of volume effects depend crucially on the $a\mu$ value: 
they are very small for large values of $a\mu$
but important when $a\mu\to 0$. As a matter of fact
for a given value of $a\mu$, there is a minimal lattice size $L$  below which the
lattice artifacts are dominant. We took this constraint into account in the present work.

\subsection{Zero modes of Wilson-Dirac Operator}
\label{spectrum}

When computing  physical observables, the integration over the fermionic fields is performed analytically and the result
is expressed  in terms of the inverse  Dirac operator  (\ref{DO}).
This has been explicitly done  in  eq. (\ref{unquenched}) for the fermion propagator which constitutes the building block of  the lattice simulations. 
When working in the quenched approximation, one can set ${\rm det}(D)=1$ and the relevant numerical task is thus reduced 
to compute $D^{-1}_{x y}[\phi]$ for an statistical ensemble of meson field configurations.
\begin{equation}
S(x,y)\ =\ \frac{1}{Z} \int[d\phi]\ D^{-1}_{xy}(\phi)\ e^{-S_{M}(\phi)}\ \approx \ \frac{1}{N} \sum_{i=1}^N D^{-1}_{xy}
(\phi_i)
\end{equation}

Due to translational invariance one is left in practice to compute $S[\phi](x,0)\equiv D^{-1}_{x 0}[\phi]$, that is to  solve the linear system:
\beq\label{LS}
D_{z x}(\phi) S_x(\phi) = \delta_{z 0}
\eeq

It is worth noticing that in the full  QFT formulation every configuration is weighted 
by the determinant of the Dirac operator $D$ and therefore the configurations yielding an
ill-conditioned linear system (\ref{LS}), i.e with ${\rm det}(D)\approx0$, do not contribute to the 
functional integral.  In the quenched  approximation, however, this is no longer true 
and ``ill-conditioned configurations'' can be sampled.

As a practical measure of the ``ill-conditioness'' of $D$  we have considered its 
``condition number'' defined as the ratio between the largest to the lowest eigenvalue 
modulus\cite{marquis}. The largest is this number the more difficult is to solve the linear system. 
Depending on the method used for that purpose, either the algorithm cannot find the 
solution, or the round-off errors make the solution wrong. In exact arithmetic the condition 
number measures how the solution changes when the second member of linear system slightly changes. 

\begin{figure}[h!]
\begin{center}
\includegraphics[width=10.cm]{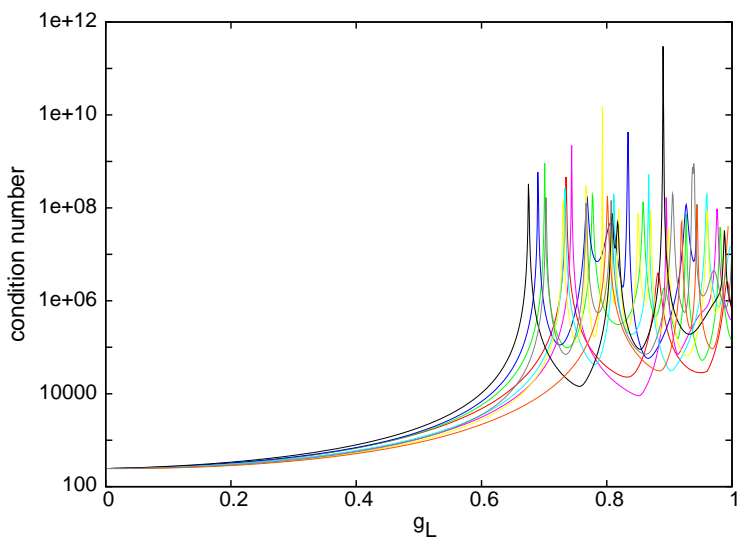}
\caption{Conditioning number as a function of $g_L$ for a fixed value of $\kappa=0.11$ and $V=8^4$ 
and for 9 different meson field configurations.}
\label{fig:cn}
\end{center}
\end{figure}

We have found that such ``ill-conditioned configurations'' appear in the Yukawa model 
for almost any $\kappa$ when $g_L \gtrsim 0.6$. In this case the inversion of the Dirac operator
becomes in practice impossible \cite{nous}.
For illustrative purposes, we have plotted in figure \ref{fig:cn}  the condition number 
of $D$ as a function of the lattice coupling constant $g_L$ for an ensemble of $L=8$ 
configurations at fixed value of $\kappa$.  As one can see,  the condition number of a given
configuration diverges on a discrete set of $g_L$ values for $g_L \gtrsim 0.6$ indicating 
the practical impossibility to compute the nucleon propagator.
The precise $g_L$ values where this divergence occurs depend on the particular 
configuration, on the values of $\kappa$ and $a\mu$  and  on the lattice size. 
It turns out however that the situation described in figure  \ref{fig:cn} is generic 
for the quenched Yukawa model. 

The existence of zero  modes in the quenched Yukawa model was already found in \cite{BDJJNS_NPB344_90} 
but the numerical results performed with very small lattice let these authors suspect
the existence of a second allowed region  at large values of $g_L$. 
It can be shown however that this result  is a volume artifact as the second region
disappears exponentially with the number of lattice sites in the thermodynamical limit. 
This issue as well as some properties of the interacting Dirac operator  (\ref{DO}), in particular the fact 
that its spectral  properties do not depend independently
of $\kappa$ and $g$ but rather on their ratio,  will be treated in detail in a 
forthcoming publication \cite{nous}.

As a conclusion,  the numerical simulations in the quenched Yukawa model
are limited to values  of the lattice coupling constant $g_L\lesssim 0.6$. 
Using a typical value of $\kappa=0.1$, this
corresponds  to $g={g_L\over2\kappa}\lesssim 3$, that is $\alpha={g^2\over4\pi}\lesssim 0.7$
which is of the same order than the $\alpha_{\rm QCD}$  in the nonperturbative region.

\subsection{Renormalized Fermion mass}

Renormalized particle masses are obtained in Lattice QFT  by considering the time evolution of
the correlator matrix defined as:
\beq\label{C}
C(\vec{p},t)\ =\ \sum_{\vec{x}} \VEV{J(x)J^\dag(0)} e^{i\vec{p}\cdot\vec{x}}\ ,
\eeq 
where $J^\dagger$ creates a particle state at the origin  and $J$ destroys it at  $x$.
The tensorial indexes -- depending on the type of particle -- are implicit and the 
vacuum expectation value $\VEV{\cdots}$ is obtained through an average over field configurations. 
It can be shown  that the correlator matrix (\ref{C}) has contributions from all the particle states $n$  satisfying  $\VEV{0|J|n}\neq 0$, and has the form
\beq
{\rm Tr }\ C(\vec{p},t)\ = \sum_n c_n  \cosh{ E_n\left( t-{T\over2} \right)   } 
\eeq 
For $\vec{p}=0$, it behaves as a sum of hyperbolic cosine with the rest mass $m_i$ of the particle states:
\beq\label{C_t}
{\rm Tr }\ C(t) &=& c_0    \cosh{ am_0\left( t-{T\over2} \right)} + c_1  \cosh{ am_1\left( t-{T\over2} \right)}  + \cdots 
\label{sum_exp}
\eeq 
Using the above equation at two consecutive times  one can extract an effective mass 
\beq\label{meff}
am_{\rm eff}(t) = {\cal F}\left[ C(t)\over C(t+1) \right]    
\eeq
which, at large enough euclidean times, will display a plateau region that will be identified 
to $am_0$ value.
 
In the case of one fermion state ($J=\Psi$),  the  correlator matrix is he $4\times4$ matrix:
\beq
C(t) &=& \sum_{\vec{x}} S(x,0)
\eeq
where $S$ is the propagator defined in Eq. \ref{unquenched}.
The case of two fermion states will be discussed in section \ref{Two_Body}.

\begin{figure}[h!]
\begin{center}
\begin{minipage}[h!]{8cm}
\hspace{-1cm}
\includegraphics[width=10cm]{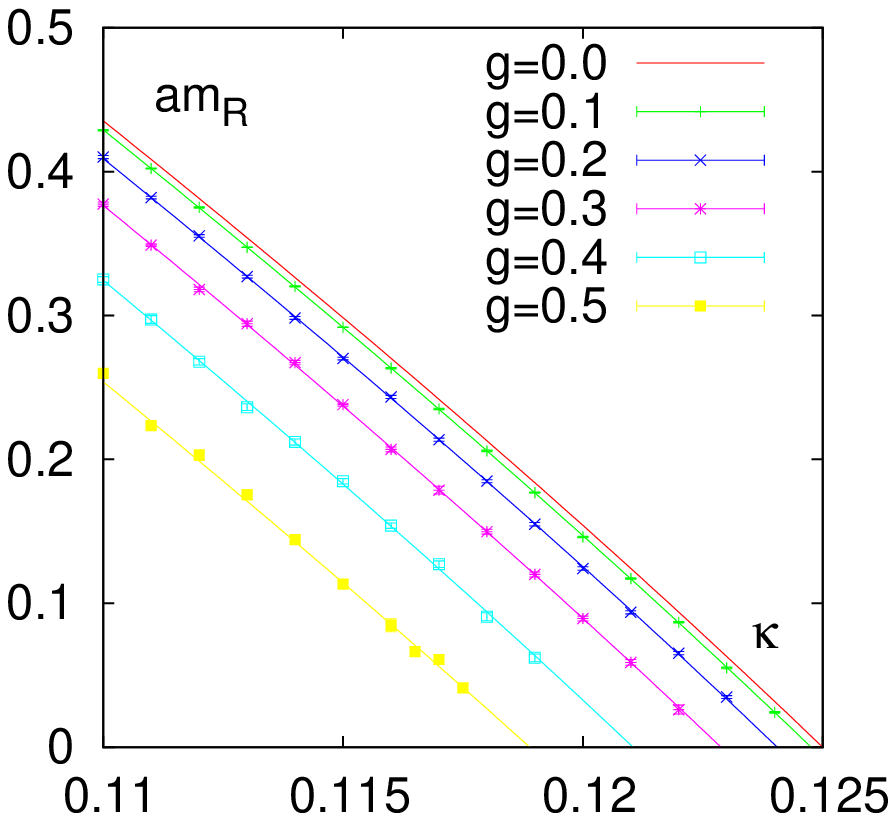}
\caption{Fermion mass {\em vs} the hopping parameter for
several values of Yukawa coupling averaged over 100 samples each one. 
The line is a best fit  according to Eq. (\ref{scaling}).}
\label{fig:m_vs_kappa}
\end{minipage}
\hspace{0.5cm}
\begin{minipage}{8cm}
\hspace{-1cm}
\includegraphics[width=10cm]{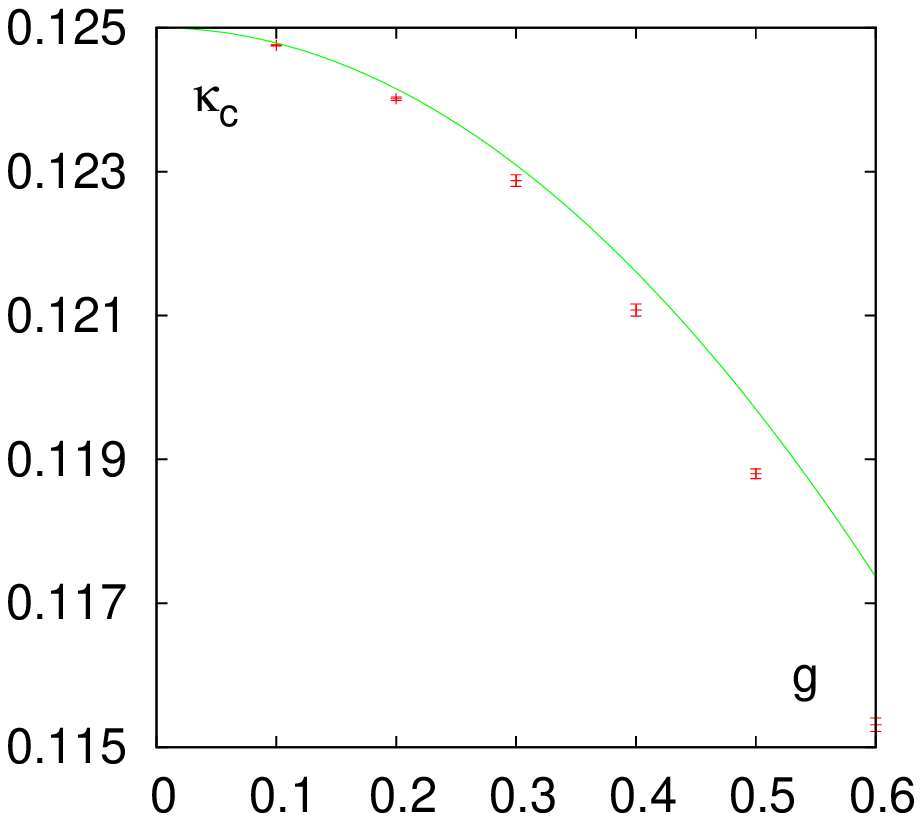}
\caption{Values of $\kappa_c$ obtained with $L=24$ and $a\mu=0.1$ 
extracted from data in figure \ref{fig:m_vs_kappa} according to Eq. (\ref{scaling}).
The line represents the perturbative behavior given by Eq. (\ref{kappac}).}
\label{fig:kappac_vs_g}
\end{minipage}
\end{center}
\end{figure}

The fermion masses  extracted in this way are presented in figure \ref{fig:m_vs_kappa} 
for $a\mu=0.1$, a lattice size  $L=24$ and several values of the
lattice coupling ranging from $g_L=0.0$ to $g_L=0.5$. 
In the free case, this mass is already different from the bare one $am_0$ due to lattice artifacts
and it is given by:
\beq
am_R = \log\left( 1 + \frac{Z}{2}\left(\frac{1}{\kappa}-\frac{1}{\kappa_c}\right)\right)
\label{scaling}
\eeq
with $Z=1$ and $\kappa_c=1/8$. 
This expression is used to fit the interacting masses in terms 
of two parameters $Z(g)$ and $\kappa_c(g)$.
The result of this parametrization is  indicated by full lines in figure 
\ref{fig:m_vs_kappa}. Note that at $\kappa=\kappa_c$ the renormalized fermion mass vanishes.
As one can see, renormalized masses are smaller with growing values of the scalar coupling constant, or in other
terms, $\kappa_c(g) < \kappa_c(0)$. This indicates that the renormalized nucleon mass is made lighter
by a scalar coupling.

The fitted coefficient $\kappa_c(g)$ is presented  in figure \ref{fig:kappac_vs_g}. 
This coefficient can be calculated in lattice perturbation theory which provides a test of numerical  simulations. 
It is quadratic in the coupling constant
\begin{equation}\label{kappac}
\kappa_c(g,L,a\mu) = \frac{1}{8}- c_2(L,a\mu) g^2 +o(g^4)
\end{equation}
with coefficient $c_2$ 
depending on the scalar mass $a\mu$ and on  the  lattice size $L$.

The region of parameter space  to perform the numerical  simulations of physical interest
is limited by the constraint $am_R>0$ which corresponds to $\kappa\in [0,\kappa_c(g,L,a\mu)]$. 
In order to avoid large lattice artifacts  the condition $am_R \ll 1$ must hold,
what restricts the useful parameter space to a narrow band in the vicinity of $\kappa_c$
with the constraint indicated in the preceding section, say $g_L\lesssim 0.6$.

\section{Two-fermion states}\label{Two_Body}

Our  main interest  in this work is to study
the interactions between fermions and thus the properties of the two-body states.
In this respect, our  reference will be the results  provided by potential models in the non relativistic ladder approximations. 
These are summarized in what follows.

\subsection{Non relativistic results}

Let us first consider the non relativistic system of two particles with equal mass $m$,
interacting  by a Yukawa potential of strength $g$ and range parameter $\mu$
\[ V(r) =- \frac{g^2}{4\pi}\; {e^{-\mu r} \over r}\]
The binding energy ($B$) and scattering length ($a_0$) are given by
\begin{eqnarray}
 {B} &=& m \;\left(\frac{\mu}{m}\right)^2 \epsilon(G) \\
 a_0 &=& {1\over\mu} \; \lambda(G) 
 \end{eqnarray}
where  $\epsilon(G)$ and $\lambda(G)$ are respectively the 
binding energy and scattering length of the dimensionless  S-wave  Schrodinger equation.
\begin{equation}\label{Sch}
u"(x) + \left[ -\epsilon + G \;{e^{-x}\over x} \right] u(x)=0  
\end{equation}
with a coupling constant $G$ related to the original parameters ($m,g,\mu$) by
\[  G = {g^2\over4\pi}\; {m\over\mu} \]

\begin{figure}[h!]
\begin{center}
\begin{minipage}[h!]{8.cm}
\includegraphics[width=7.cm]{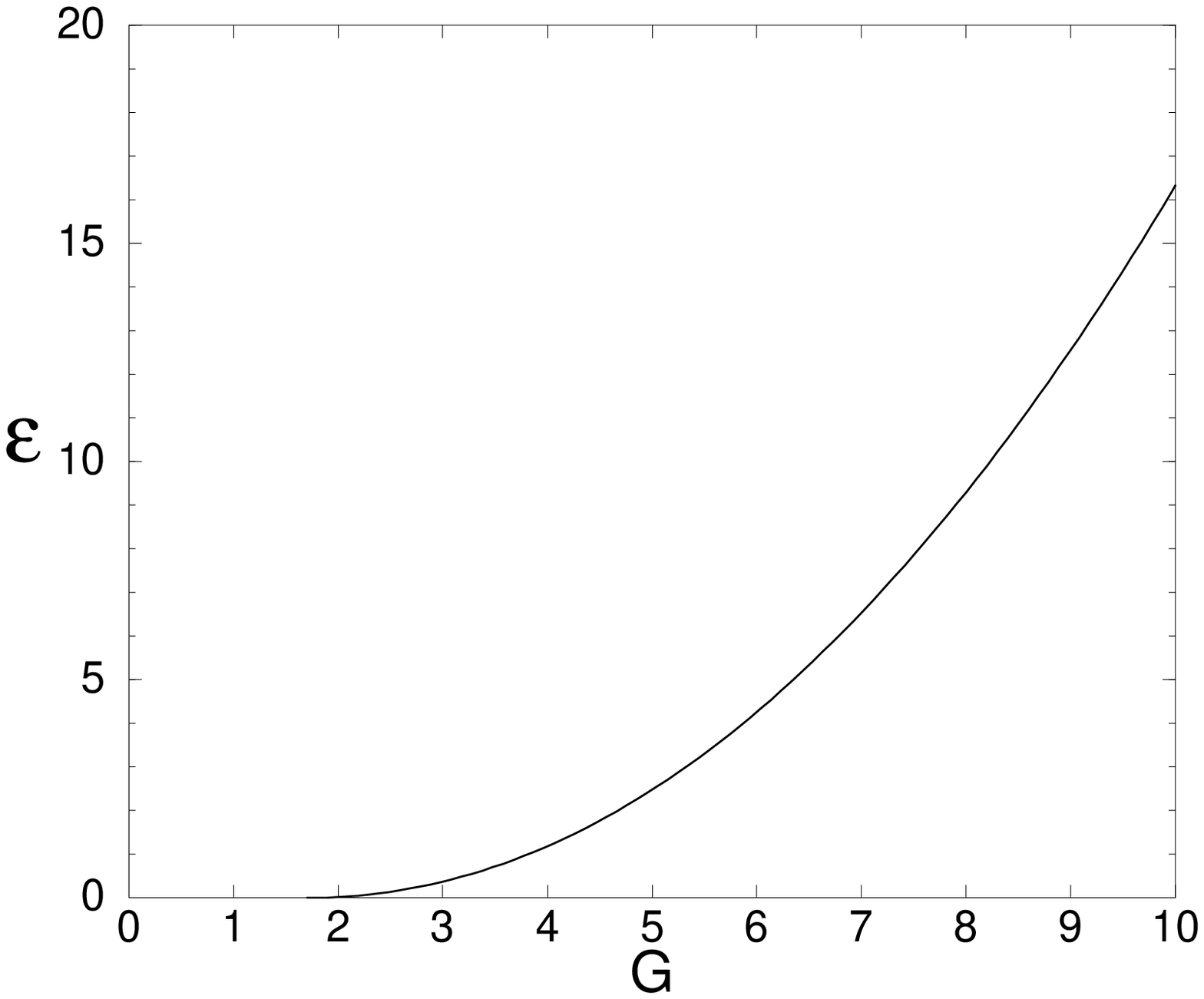}
\caption{Ground state binding energy of the dimensionless non relativistic  Yukawa model (\ref{Sch}) as a function of the coupling constant $G$. 
The appearance of the first bound state corresponds to $G_0=1.680$.}\label{eps_G}
\end{minipage}
\hspace{0.5cm}
\begin{minipage}[h!]{8.cm}
\includegraphics[width=7.cm]{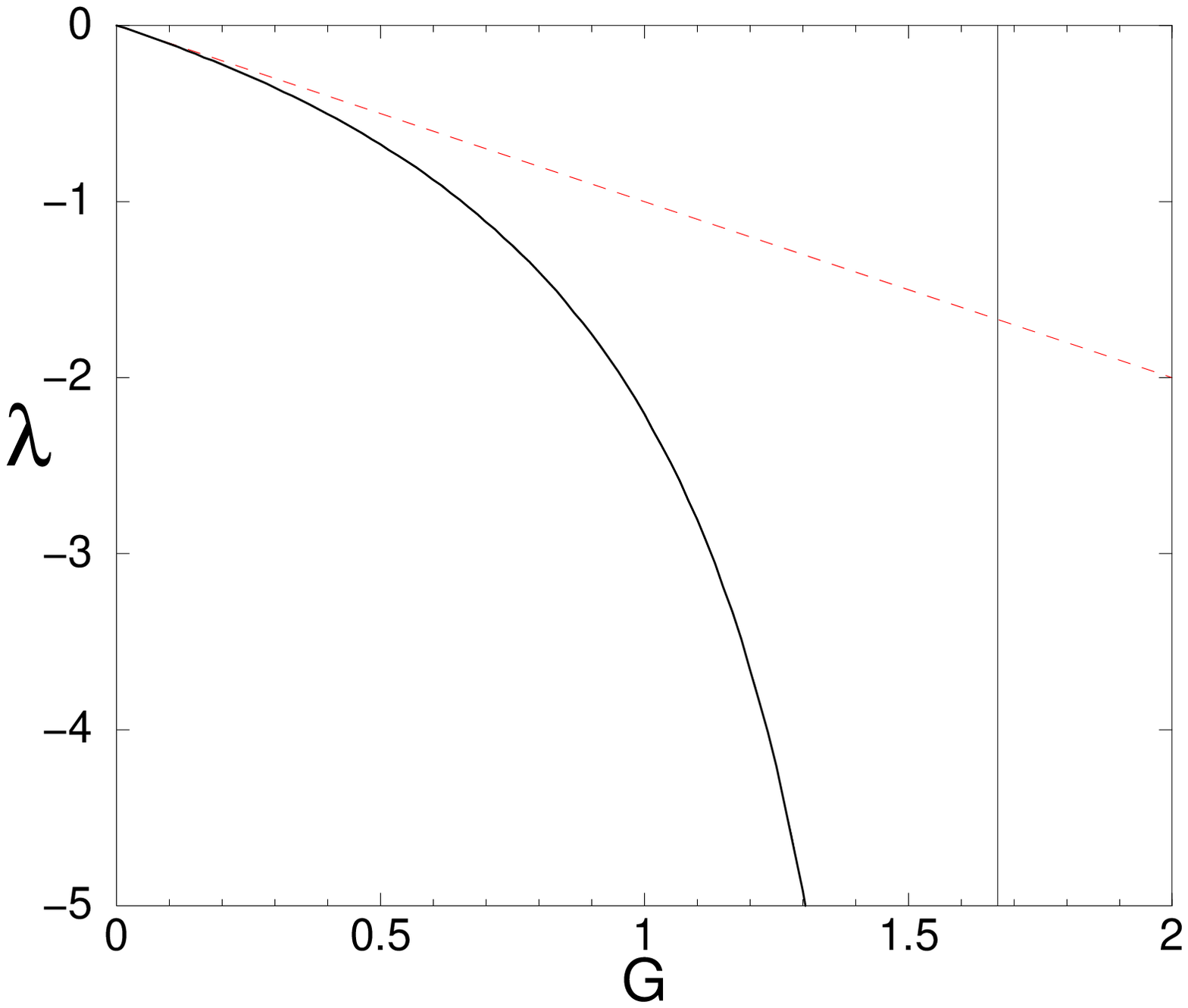}
\caption{Scattering length in the dimensionless non relativistic Yukawa model as a function of the coupling constant $G$ (solid line).
The Born approximation is indicated by the dashed line. 
The singularity corresponds to the appearance of the first bound state.}\label{lambda_G}
\end{minipage}
\end{center}
\end{figure}

The functions $\epsilon(G)$ and $\lambda(G)$
are displayed in figures \ref{eps_G} and  \ref{lambda_G}.
The convention used for the scattering length corresponds to 
$\delta(k)=-a_0k +o(k^2)$.
The critical value for the appearance of the ground state is $G_0\approx1.680$. 
At this value $\lambda(G)$ has a pole and it can be shown that for small  values of G one has 
\begin{equation}\label{Born} 
\lambda(G) = -G +o(G^2)
\end{equation}
which corresponds to the Born approximation.

\subsection{Binding energies}

We will restrict ourselves  to  study the system of two identical fermions, say $NN$, in the $J^{\pi}=0^+$ state. 
The general form of the interpolating field $J(x)$ for a two fermion state  reads
\beq\label{J2}  
J(x)=  \sum_{\alpha\beta} \psi_{\alpha}(x)\Gamma_{\alpha\beta}\psi_{\beta}(x)
\eeq
where $\Gamma$  depends on the quantum number of the
state.  For a $0^+$ state one has $\Gamma=i\gamma_2\gamma_0\gamma_5$. In the matrix form (\ref{J2}) can be written as 
\[
J(x)= \psi^t_x \Gamma \psi_x 
\]
Using the interpolating field (\ref{J2}), the  $NN$ correlator becomes 
\[
C(t)=\sum_{\vec{x}} \langle0|J(0)J^{\dagger}(x)|0\rangle 
    =\sum_{\vec{x}} \langle0|\psi_0\Gamma\psi_0\bar{\psi}_x\gamma_0\Gamma\gamma_0 \bar{\psi}_x|0\rangle 
    =2\ \rm{Tr}\sum_{\vec{x}}  \langle 0|\Gamma S(x,0)\Gamma S(x,0) |0\rangle 
\]
where $S(x,0)$ denotes the fermion propagator.

In order to decrease the contribution
of the excited states we have introduced a smearing procedure. 
This consists  in  modifying the interpolating field $J(x)$ in the following way:
\begin{equation}
J(x) = \sum_{\vec{R}} f(\vec{R})  \psi_{\vec{x},t}\Gamma\psi_{\vec{x}+\vec{R},t}
\end{equation}
where $f(\vec{R})$ is some smearing function to account
for the spatial extension of the state. As we are interested in the s-wave, 
the smearing can be done over the whole timeslice in an efficient and inexpensive way by 
choosing the smearing function  to be  a constant, $f(\vec{R})=1/\sqrt{L^3}$. 
After performing the appropriate Wick contractions, the time-correlator results:
\beq \label{C2smeared2}
C(t) &=&  \VEV{\Gamma \widetilde{C}_1(t) \Gamma {C}^t_1(t) - \Gamma C_1(t) \Gamma^t \widetilde{C}^t_1(t)}
\eeq
where the 4x4 matrices $C_1(t)$ and $\widetilde{C}_1(t)$ are respectively:
\beq
C_1(t) &=& \sum_{\forall \vec{x}}S(\vec{x},t) \\
\widetilde{C}_1(t) &=& \sum_{\forall \vec{x}}\widetilde{S}(\vec{x},t) 
\eeq
and $\widetilde{S}(\vec{x},t)$ is the solution of the linear system
\beq
D^{\alpha\beta}_{y x} \widetilde{S}^{\beta\gamma}_{x}  = \frac{1}{L^3}\sum_{\vec{R}}  
\delta^{\alpha\gamma} \delta_{\vec{y} \vec{R}} \delta_{y_0 0} \ .
\eeq

\begin{figure}[h!]
\begin{center}
\begin{minipage}[h!]{8cm}
\includegraphics[width=8.cm]{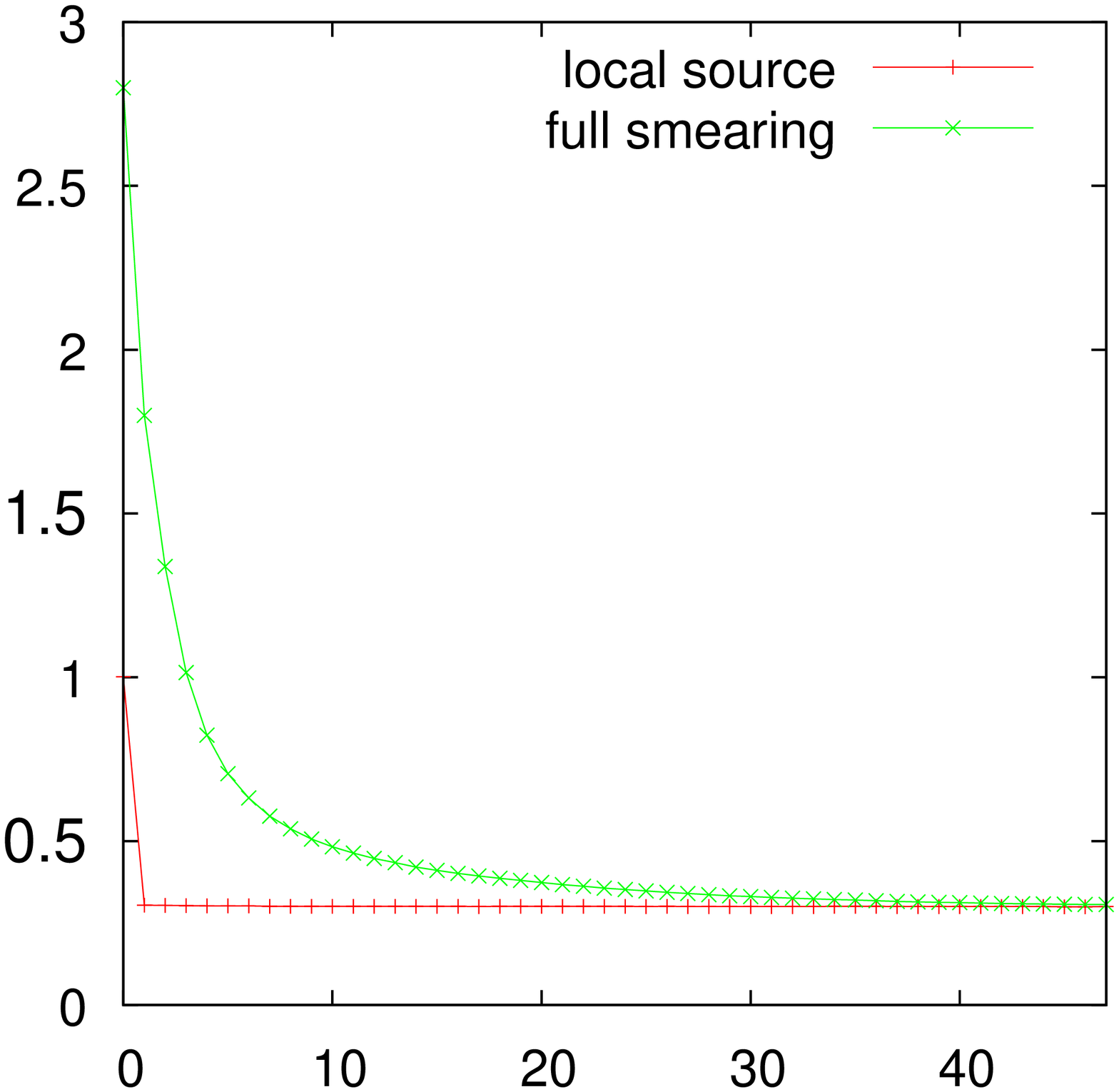}
\caption{Two-body effective mass for a two-fermion $0^+$ 
state {\it vs} time with local source and fully smeared sources.
$L=48$, averaged over 800 configurations, $g_L=0.3$, $\kappa=0.118$, 
and $a\mu=0.1$.} 
\label{fig:meff2}
\end{minipage}
\hspace{0.5cm}
\begin{minipage}[h!]{8.cm}
\begin{center}
\includegraphics[width=8.cm]{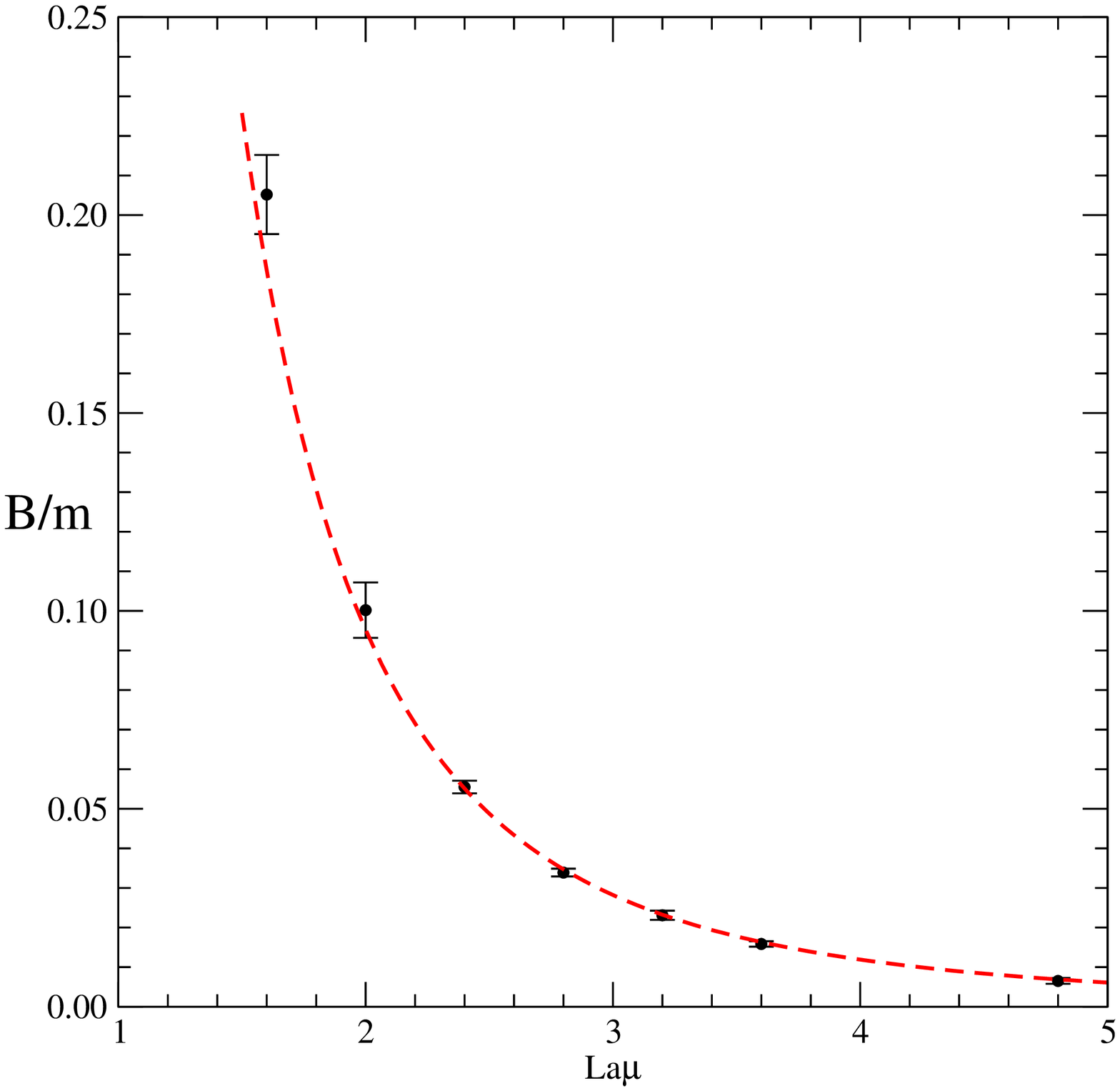}
\caption{Binding energy versus lattice size $L$, for $g_L=0.3$, 
$\kappa=0.118$, and $a\mu=0.1$ averaged over 4000 samples for $L=16,\cdots,32$, 2000 for $L=36$, 
and 800 for $L=48$. Dotted line corresponds to a $1/L^3$ fit. }
\label{fig:B_vs_L}
\end{center}
\end{minipage}
\end{center}
\end{figure}

This three-dimensional smearing  efficiently removes higher energy contributions to the correlator as it can be seen in 
figure \ref{fig:meff2} where the effective mass (\ref{meff})
is plotted both for the local and smeared interpolating fields.
It can be seen that  for large $t$ values the effective mass tends 
to a plateau which actually defines  the mass of the state. 
If the local sources were used we would have needed larger values of 
$t$ to be able to  find a plateau in the masses. 

An interesting property of this smearing is that the two-body  free correlator is 
the square of the one-body one at any value of $t$, i.e.,
the effective mass of the two-body state is rigorously 
constant and exactly twice that of one particle. 

The binding energy of a two-fermion  state with total mass $m_2$ is defined as $B=2m_1-m_{2}$ where $m_1$ denotes the mass of the fermion. 
In figure \ref{fig:B_vs_L} we show this binding energy
as a function of the lattice size $La\mu$ for a given set of parameters. The dotted line is a fit obtained with a $1/L^3$ dependence.
As it can be
seen in this figure, the binding tends to zero in the infinite volume limit. This indicates that this two-fermion system has no bound state  
for this particular set of parameters. The situation is however the same for the whole range
of parameters  accessible in the numerical simulations. 
Although assuming that this  pathology  could be associated 
to the quenched approximation it is physically surprising  that no any NN bound state could be generated  
unless the N\={N} pairs  creation (unquenched Yukawa model) are taken into account.

\begin{figure}[h!]
\begin{center}
\begin{minipage}[h!]{8.cm}
\begin{center}
\includegraphics[width=8.cm]{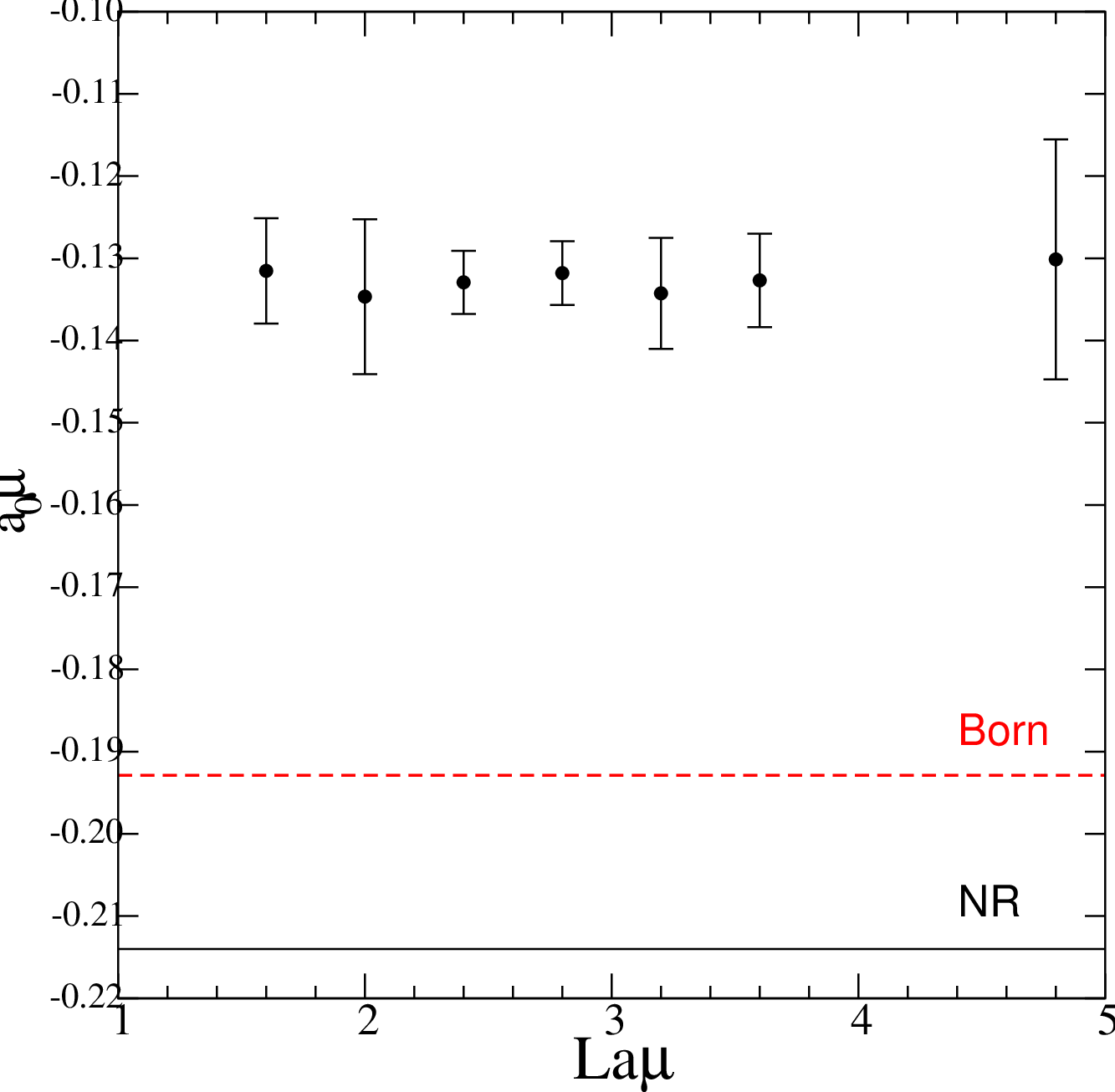}
\caption{Scattering length extracted from Eq. (\ref{a0naive}) 
as a function of the lattice volume $La\mu$,  for $g_L=0.3$, 
$\kappa=0.118$ and $a\mu=0.1$ {\it i.e.} $G=0.193$, extracted 
from the data in fig. \ref{fig:B_vs_L}. 
The solid line indicates the non relativistic
results and the dotted  the Born approximation (\ref{Born}).}\label{fig:a0mu_vs_L}
\end{center}
\end{minipage}
\hspace{0.5cm}
\begin{minipage}[h!]{8.cm}
\begin{center}
\includegraphics[width=8.cm]{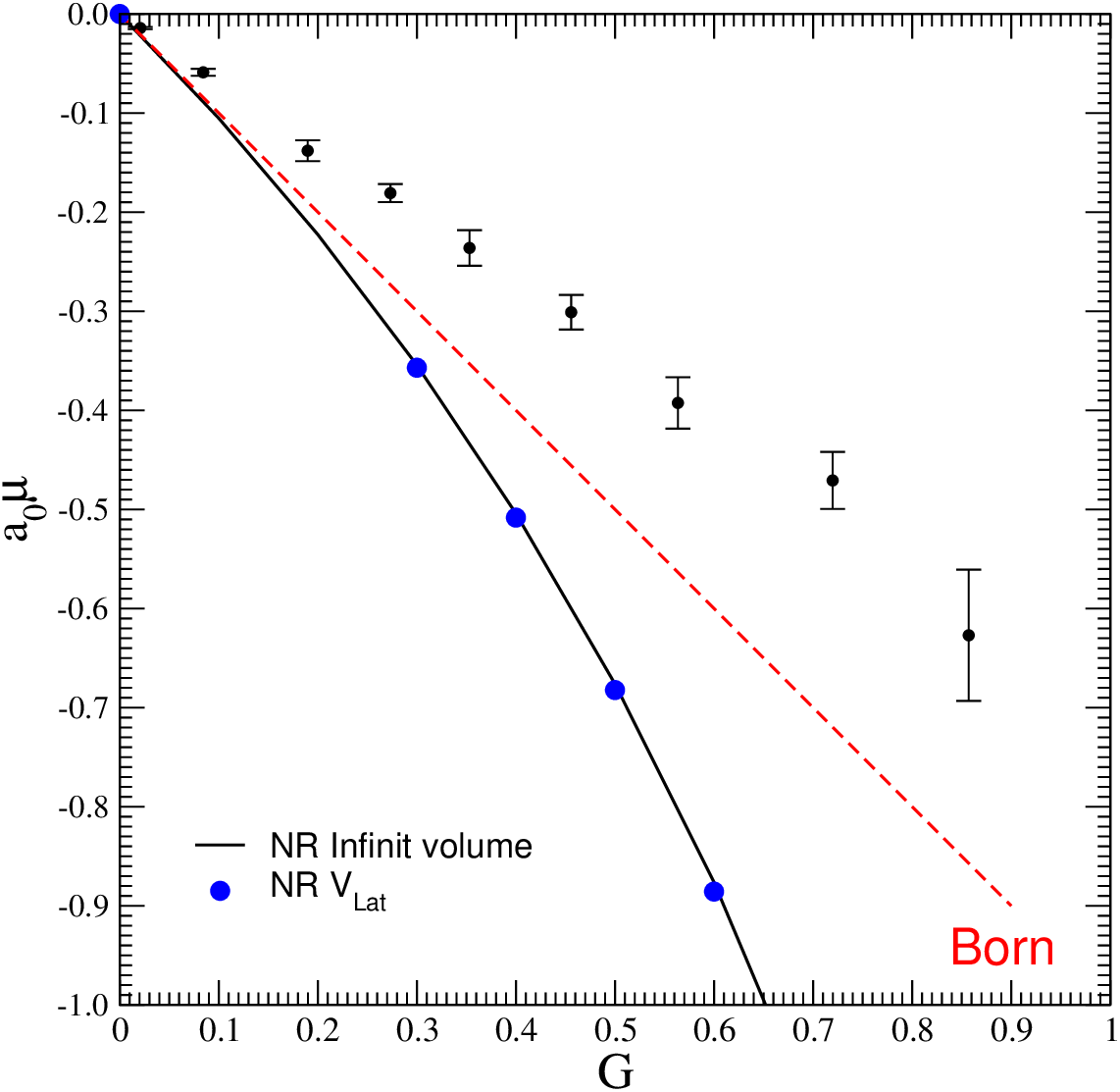}
\caption{Scattering length vs $G$ for a 
lattice volume $La\mu=2.4$, 2000 samples for each point up to $G=0.3$ and $5000$ thereafter.
The solid line indicates the continuum non-relativistic
result and the dotted one the Born approximation (\ref{Born}).
Blue circles represent the NR result obtained using the lattice potential in figure 
\ref{latticepotential}.}
\label{fig:a0_naive}
\end{center}
\end{minipage}
\end{center}
\end{figure}

\subsection{Scattering lengths}\label{sec:LECs}

Since there is no bound state below the critical coupling constant, we can only access
to the scattering properties of the state. The scattering observables cannot be obtained in Euclidean
time in the infinite volume limit \cite{MT_NPB_245_90} but can be extracted from  the volume dependent binding energy  measured on
finite lattices, like for instance the one plotted in figure \ref{fig:B_vs_L}. 
The underlying formalism was developed by Luscher in \cite{ML_CMP104_86,ML_CMP105_86} who gave a $1/L$ expansion of the
the binding energy . In its leading order  it reads: 
\begin{equation}
\frac{B}{m} = - \frac{4\pi a_0\mu}{\left(\frac{m}{\mu}\right)^2 (La\mu)^3}  \label{a0naive}
\end{equation}

Taking the binding energy values of figure \ref{fig:B_vs_L} and equation 
(\ref{a0naive}), the NN scattering lengths $a_0$ have been extracted.
The results corresponding to  $g_L=0.3$, 
$\kappa=0.118$, and $a\mu=0.1$  are displayed in fig \ref{fig:a0mu_vs_L}. They manifest a constant behavior of $a_0$ as a function of the
lattice size L, indicating that the first order expression we used  to extract the scattering length is justified.
The dimensionless coupling constant of the nonrelativistc model for the parameters of figure \ref{fig:a0mu_vs_L} is $G=0.193$.
The corresponding non relativistic scattering length value, given by figure \ref{lambda_G},  is $A_0=a_0\mu=-0.214$ (solid line), quite close to its Born approximation (dashed line).

This study has been performed for several values of $g_L$. The dependence 
of $a_0$ on the coupling constant $G$ is plotted in figure \ref{fig:a0_naive},
for a lattice size of $La\mu=2.4$ ($L=24$,  $a\mu=0.1$).
One can see that the lattice results notably departs
from the non relativisitc ones (solid line) and are above the Born approximation (dashed line).  
In figure \ref{fig:a0_naive} the NR scattering length has been computed using the lattice
discretized potential of figure \ref{latticepotential}. 
As one can see, the regularization of the potential at the origin has no effect
in the scattering length (indistinguishable from the continuum result in the figure). 
This is due to the fact that the scattering length is a zero energy 
observable and therefore is not 
very sensitive to the details of the interaction.
From the preceding analysis we conclude that 
the repulsive effect shown by the lattice data is not related to the 
lattice potential discretization at short distances.

The values of the accessible coupling constants
extend beyond  the Born regime but are still far from the pole
behavior corresponding to the appearance of the first bound state displayed in figure \ref{lambda_G}.
The difference between the lattice and NR results may indicate strong repulsive corrections.
These kind of corrections are already manifested in the bound state problem
when solving the same Yukawa model both in 
Light Front \cite{MCK_PRC68_2003} and  Bethe-Salpeter \cite{CK_EPJA_2010} ladder equations.

\section{Conclusion}\label{Conclusion}

We have considered the quantum field theory solution of the simplest nuclear Yukawa model
consisting on two identical nucleons interacting via a scalar meson exchange.
The choice of the scalar coupling with respect to pseudo scalar one 
was taken in order to optimize the appearance of two-body bound states we were interested in.

The problem has been solved using the standard lattice techniques, based on the path integral formulation
of the theory on a discretized space-time.
The meson field has been described by  a discretized Klein-Gordon Lagrangian  without self-interacting $\lambda\phi^4$ term
and the Dirac-Wilson discretization was chosen for the fermion.

The resulting model is  fully relativistic and was solved  by neglecting only the N\={N}
loops generated from the meson field in the so called quenched approximation.  This simplification is
physically justified by the heaviness of the nucleon and is anyway implicit  in all nuclear models.

The numerical simulations were performed along the physical line $\mu/m\approx0.6$
where $\mu$ and $m$ denote respectively the meson and nucleon renormalized masses.
The solutions were found only for coupling constants below
some critical value $g\lesssim 3$. Above this value the ubiquitous presence of
fermion zero modes made the problem numerically unsolvable.
The addition of a pseudo scalar coupling term does not make the problem simpler.
The present situation does not allow to judge whether this problem is related to 
the particular fermion discretization used. However the same problem
was described in the past to affect naive fermions \cite{Yukawa}. This seems to indicate 
that the use of Wilson fermions is not responsible for the problem.

The range of the accessible coupling constants is below the 
threshold value for producing two-body bound states, which in the non relativistic potential
approximation turns to be $g\approx 3.7$ and in the Bethe-Salpeter one sensibly larger.
In the accessible region of $g$ the NN scattering length was calculated using the Luscher procedure.
The values found were in agreement with the non relativistic models
for low coupling constants but show strong repulsive effects when increasing $g$.

We conclude that the  quenched approximation of the Yukawa model (scalar coupling)
is not able to produce two-nucleon bound states. Although this pathology is manifested
in the quenched approximation it suggests that the non-relativistic results 
based on one-boson exchange  potentials have no direct counterpart in the quantum 
field theory approach.

\section*{Acknowledgment}

The authors are pleased to acknowledge the fruitful discussions held at LPT with Ph. Boucaud, J.P.
Leroy and O. Pene who fully participated in the early stages of this work. We thank the staff of the
Centre de Calcul IN2P3 in Lyon where some of the numerical  calculations were performed. This works
has benefit from the French-Spanish Collaboration Agreement IN2P3- MICINN. 


\end{document}